\documentclass{aa}
\usepackage{graphicx}
\begin{document}
   \title{The $r'$-band luminosity function of Abell~1367: a comparison with Coma   \thanks{Table~2 and catalogs described in Appendix are only available in electronic form at the CDS via anonymous ftp to cdsarc.u-strasbg.fr}
}


   \author{J. Iglesias-P\'{a}ramo
          \inst{1}
          \and
A. Boselli
          \inst{1}
          \and
   G. Gavazzi
          \inst{2}
          \and
   L. Cortese
          \inst{2}
          \and
   J.M. V\'{\i}lchez
          \inst{3}
          }

   \offprints{J. Iglesias-P\'{a}ramo}

   \institute{
Laboratoire d'Astrophysique de Marseille, BP8, Traverse du Siphon, 
F-13376 Marseille, France\\
              \email{jorge.iglesias@astrsp-mrs.fr,alessandro.boselli@astrsp-mrs.fr}
         \and
	Universit\'{a} degli Studi di Milano, Bicocca, Piazza delle Scienze, 3,
20126 Milano, Italy\\
	\email{gavazzi@mib.infn.it,luca.cortese@mib.infn.it}
	\and
             Instituto de Astrof\'{\i}sica de Andaluc\'{\i}a (CSIC), Apdo. 3004, 18080 Granada, Spain\\
             \email{jvm@iaa.es}
            }

   \date{Received 4 June 2002; accepted 10 October 2002}

   \abstract{
We made a large (approximately $1\degr \times 1\degr$) 
$r'$-band imaging survey of the central regions of the two nearby clusters of
galaxies, Abell~1367 and Coma. The data, presented as a catalog, are used to
construct the $r'$-band luminosity function (LF) of galaxies in these
two clusters, by subtracting the Yasuda et al. (2001) galaxy
counts from our cluster counts. 
Our Coma luminosity function is consistent with previous determinations, 
i.e. providing a faint end slope $\alpha = -1.47_{-0.09}^{+0.08}$, 
significantly steeper than 
the one we find for Abell~1367 ($\alpha = -1.07_{-0.16}^{+0.20}$). 
The counts in Abell~1367 show a relative minimum at $r' \approx 19$, followed by a steep 
increase faintward.
The difference between the two clusters appears significant, given the consistency of 
the experimental conditions in the two clusters.
Whereas for Coma we find a significant increase 
of the slope of the LF outwards, no 
such effect is found for Abell~1367.
   \keywords{atlases --
              galaxies: general   --
	galaxies: clusters: general
               }
   }

   \maketitle
%

\section{Introduction}

The study of the LF of galaxies provides us with a diagnostical tool indispensable 
for solving one of the hottest, yet unsettled cosmological issues, i.e. a plausible 
reconstruction of the evolution of galaxies from the epoch of their formation to the present.
The comparison of the LF of galaxies in clusters and in the field, for example, 
should shed light on the role of the environment in regulating the evolution of 
galaxies, both for the giant and dwarf populations 
(see Press \& Schechter 1974, Binggeli et al. 1988).
Moreover the comparison of the galaxy LF in nearby clusters with that of 
clusters at progressively larger $z$ can improve our knowledge of the evolution of 
galaxies in a given environmental condition.
The recent work by Trentham \& Tully (2002) on the LF  in five
different local environments up to $M_{R} = -10$, has shown that there are
far fewer dwarfs than what expected from CDM models.
The Coma cluster, being the prototype of evolved rich
nearby clusters, is the one on which the attention has been most focused. 
After the catalogue of galaxies in the Coma cluster (Godwin et al. 1983),
several studies on this cluster have been published
in different wavelength windows 
(e.g. Donas et al. 1991; Biviano et al. 1995;
Lobo et al. 1997; Andreon 1999; Andreon \& Pell\'{o} 2000). 
Considering the optical $R$-band alone, there is little agreement in the 
literature on the faint-end slope of the LF. The published values span 
from $\alpha \approx -1.20$ (Beijersbergen et al. 2002) to $\alpha \approx -1.7$ (Trentham 1998). 
Intermediate values of $\alpha
\approx -1.30$ (Lugger 1989) and $\alpha \approx -1.40$ (L\'{o}pez-Cruz et al. 1997; 
Bernstein et al. 1995; Secker 1997; Thompson \& Gregory 1993) are also found. 
This remarkable lack of agreement is probably due to the
inhomogeneity of the observational and data processing conditions,
and
different magnitude ranges or different cluster regions contributing to the determination
of the LF.\\
Abell~1367, conversely, in spite of its many interesting structural aspects, 
has received little attention so far.
After the first catalog of galaxies from Godwin \& Peach (1982) very few studies have
been carried out on the determination of the LF. Although Trentham
(1997), attempted a
LF in the $R$ band, he could not perform a quantitative study of the
LF because of the large uncertainties in the 
background subtraction. On the other hand Schechter (1976) and Kashikawa et
al. (1995) presented fittings of the LF of Abell 1367 to analytical
functions but they kept
the parameter $\alpha$ fixed to $-1.25$. Lugger (1986) obtained a value of
$\alpha = -1.42$ for the LF of Abell~1367 but his limiting magnitude
was not deep enough to include
the dwarf galaxies.
In this work we present the 
$r'$-band 
LF of Abell~1367 that includes the dwarfs, and we compare it with the one of Coma,
obtained under the same observational and instrumental conditions.
Since both clusters have approximately the same recessional velocity (7000 and
6500~km~sec$^{-1}$ respectively for Coma and Abell~1367), the limiting
magnitudes as well as the selection biases are comparable.
The differences between the two LFs, if any, should then be interpreted as due to
the different evolutionary histories of the two clusters or to different initial
conditions rather than to other observational or instrumental circumstances. 
A value of $H_{0} = 75$~km~s$^{-1}$~Mpc$^{-1}$ is adopted
throughout this paper.

The paper is arranged as follows: Section~2 contains the details on the
observations. Section~3 and 4 give details on the
extraction of sources and the photometric calibration. 
Section~5 lists the different entries contained in the catalogs.
Section~6 shows the LFs for both clusters and section~7 contains a brief
summary and discussion of the results. A description of the catalogs 
is presented in Appendix~A.

\section{Observations}

The data presented in this work are a by-product of an H$\alpha$ survey of
nearby clusters aimed at constructing their H$\alpha$ LFs, whose main results
can be found in Iglesias-P\'{a}ramo et al. (2002).
Observations were carried out with the Wide Field Camera (WFC) attached to
the Prime Focus of the INT~2.5m located at Observatorio de El Roque de
los Muchachos, on April 26th and 28th 2000, under photometric
conditions, excepting the last half of the second night. 
However, since several exposures of each field were
taken, we could properly calibrate all the data. The average seeing ranged from
1.5 to 2~arcsecs on both nights.

The WFC for the INT comprises a science array of four thinned AR
coated EEV 4K$\times$2K devices, plus a fifth acting as
autoguider. The pixel scale at the detectors is 0.333
arcsec~pixel$^{-1}$, which gives a total field of view of about $34\times 34$~
arcmin$^{2}$. Given the particular arrangement of the detectors, a squared area
of about $11 \times 11$~arcmin$^{2}$ is lost at the top right corner of the field.
At the end, four fields covering a surface of about $1\degr \times 1\degr$ were observed 
(see Figure~2 in Iglesias-P\'{a}ramo et al. 2002). The total exposure time for each field
was $3 \times 300$~s, except for one field in Coma for which
a single 300~s exposure was taken.

No broad band photometric standards were observed since the observational run
was not devoted to produce broad band catalogs. However, relative calibration
between the different frames were possible given that spectrophotometric
standards for H$\alpha$ calibration were observed.
Detailed information about the observations and data reduction procedure can be
found in Iglesias-P\'{a}ramo et al. (2002).

\section{Extraction of Sources}

The identification and extraction of sources was carried out using the
code {\em Sextractor} (see Bertin \& Arnouts 1996, for details). The limiting
size and limiting flux for detection of objects were set to 49 pixels (which
corresponds approximately to the size of the seeing disk in the frames)
and 2.5 times the standard deviation of the sky, respectively, in
order to minimize the number of spurious detections. 

The regions corresponding to the wings of the PSF of bright, saturated stars were
removed. 
The separation between stars and galaxies is based on the G/S
parameter, $\rho$, 
given by {\em Sextractor} (see Bertin \& Arnouts 1996). 
To the first approximation, we accept as
galaxies all objects with $\rho < 0.05$, and stars those with $\rho >
0.95$. For the rest of the objects, a closer
inspection based on the FWHM estimated by the IMEXAM task running on
IRAF\footnote{IRAF is distributed by the National Optical Astronomy Observatories,
which are operated by the Association of Universities for Research
in Astronomy, Inc., under cooperative agreement with the National
Science Foundation.} was performed and those which were undoubtedly found
to be stars were removed. We point out the possibility of
loosing compact galaxies that are unresolved in our CCD frames in the
process of rejecting stars as suggested by Andreon \& Cuillandre
(2002). In order to test whether a population of compact dwarf
galaxies was lost in the inner parts of the clusters, we performed counts of our
rejected unresolved objects with magnitudes in the range $-18 \leq
R_{C} \leq -15$\footnote{this is the range of magnitudes found for the
BCDs} 
over the inner disk $D \leq 0.4$~degrees
and the outer annulus $D \geq 0.5$~degrees. Surprisingly we found an
excess of counts per squared degree in the outer annulus for
Abell~1367 whereas for Coma the excess of counts per squared degree
was found in the inner disk. Thus, no conclusive results of a
systematic lost of compact galaxies can be stated from our data.

An astrometric solution was found with the USNO\footnote{United States Naval
Observatory} catalog of stars for each individual frame. 
After checking the rms of the fitting and the absolute offsets between the
coordinates found for those objects appearing in more than one frame, 
the accuracy of this solution was found to be
approximately 2~arcsecs. 

A total of 149 galaxies 
common to more than one frame was removed by direct inspection. When multiple
detections existed for an object, we ruled out the one presenting the worst quality. The
criteria to decide the quality of the detections include the distance to the border of the
frame, the existence of halos of diffuse light from bright stars and  the vignetting of the
North part in detector \#3. The measured
photometry for the multiple
objects was also used to obtain the relative calibration between the different
detectors. 
\section{Photometry}

The final instrumental magnitudes are derived using {\em Sextractor}'s MAG\_BEST
magnitudes. A total magnitude is provided for the galaxies following Kron's (1980) first
moment algorithm. However for those galaxies suspected to have a neighbor biasing the
magnitude by more than 0.1~mag, a corrected isophotal magnitude is provided (see Bertin \&
Arnouts 1996 for more details).
For the objects for which the corrected isophotal magnitude was
obtained, we compared it with that computed by performing detailed aperture photometry.
In Table~\ref{close} we show the fraction of galaxies for which our aperture
magnitude differs by more than 0.2~magnitude from the magnitude provided by {\em
Sextractor}, in three magnitude intervals. 
Even for most of the faintest galaxies, the {\em
Sextractor}'s magnitudes agree with our more accurate 
aperture magnitudes within 20\%, which is
accurate enough for a statistical analysis of the LF.

The instrumental zero point was first estimated from the H$\alpha$ 
calibration, taking into 
account the airmass, the relative calibration between the different detectors and the 
scaling factor between the broad band and the H$\alpha$ frames, which were taken
respectively with a Sloan-Gunn $r'$ and a narrow band [S{\sc ii}] filters (see
Iglesias-P\'{a}ramo et al. 2002 for details on the instrumental
setup). This provided us with instrumental magnitudes already
corrected for airmass and for chip-to-chip variation, that could be
directly compared with the published photometry.

The procedure used for the calibration of our data is the following: we
searched in the literature for all galaxies belonging to any of the two
clusters with available aperture photometry in any photometric band with an
effective wavelength close to 6000~\AA, namely $R$ Cousins, $r$ Gunn and $r'$ Sloan.
The relationship between our instrumental magnitudes ($M_{ins}$) and the ones corresponding
to the different photometric systems at similar effective wavelengths are:
\begin{equation}
R_{Cousins} = m_{ins} + Z_{Cousins}
\label{johnson}
\end{equation}
\begin{equation}
r_{Gunn} = m_{ins} + Z_{Gunn}
\label{gunn}
\end{equation}
\begin{equation}
r'_{SDSS} = m_{ins} + Z_{SDSS}
\label{sdss}
\end{equation}
respectively for the Cousins, Gunn and Sloan photometric systems.

Expected values for the differential zero points are
\begin{equation}
Z_{Gunn} - Z_{Cousins} = 0.35
\end{equation}
and 
\begin{equation}
Z_{Gunn} - Z_{SDSS} = 0.15
\end{equation}
from Jorgensen (1994) and Fukugita et al. (1996).

We did not apply a color correction because of the lack of any color
information for the target galaxies. Given the similarity of the filter profiles
($\lambda_{eff}$, $\Delta \lambda$) in the three photometric systems mentioned
above, we expect an almost negligible color term in the conversion equations,
thus our instrumental magnitudes can be converted to any of the three systems
(see Fukugita et al. 1995 for details on the color effects for the different
galaxy types in different photometric systems). Furthermore, if any color term
is present, it would be negligible compared to the bin width used in the
construction of the LF ($\Delta r' = 0.5$~mag), thus unaffecting the main
results of the present study.

Jorgensen et al. (1992) give Gunn $r$-band aperture photometry
for a large
sample of ellipticals and S0s belonging to the Coma cluster. We performed
instrumental photometry of the galaxies in common using the same apertures. 
For cross-calibration we used the galaxies with intermediate magnitudes 
(in the range $14.0 \leq r' \leq 16.0$). The faintest galaxies were excluded 
because they have the largest statistical errors. 
Marin-Franch \& Aparicio (2002) performed Cousins $R_{C}$-band aperture photometry
for a sample of galaxies in the central regions of Coma. Their data were taken
at the 2.5m INT, with the same instrumental configuration as ours. They made 
aperture photometry of eight galaxies in common with our catalog. 

\begin{figure*}[t]
   \centering
    \includegraphics[width=8.5cm]{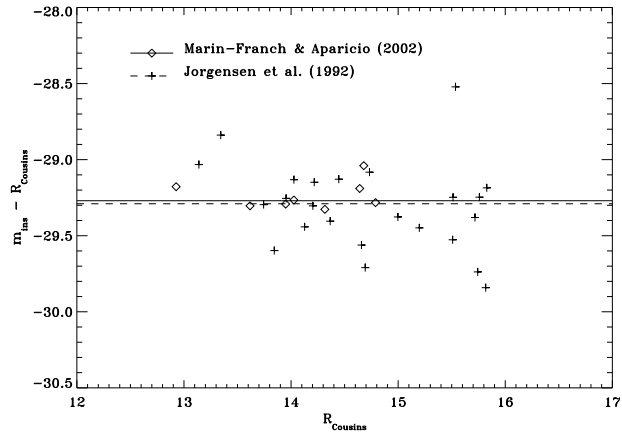}
      \caption{
$R$ magnitude vs. the difference between our instrumental magnitudes and $R$-band
magnitudes from Jorgensen et al. (1992, crosses) and Marin-Franch \& Aparicio (2002, open
diamonds). Both sets of data are expressed in Cousins $R$ magnitudes. The horizontal lines
represent the median values of both distributions.
}
         \label{compa_jorg}
   \end{figure*}

Figure~\ref{compa_jorg} shows the difference between our
instrumental magnitudes and the published magnitudes vs. the $R_{C}$
magnitudes. 
The horizontal lines show the median
value of the distributions. The best fitting results for the zero points expressed in
Cousins $R$ magnitudes are $Z_{Gunn} = -28.90 \pm 0.15$ (for the data of Jorgensen et
al. 1992) and $Z_{Cousins} = -29.27 \pm 0.09$ (for the data of Marin-Franch \& Aparicio
2002). 

After correlating both independent calibrations, we obtained $Z_{Gunn} -
Z_{Cousins} = 0.37$, which is in good agreement with the expected value for
these two photometric systems (Eq.~4). For Abell~1367 we adopt the same value of
$Z_{Cousins}$ since the night was photometric and the same instrumental setup
was used.

In addition we applied average corrections for Galactic extinction
following Schlegel et al. (1998), that
amounts to 0.02 and 0.06~mag for Coma and Abell~1367 respectively.

In what follows, we will refer our photometry to the Sloan $r'$-band system
for convenience (since the used background counts are refereed to this
system, see section~6 for more details) using Eqs.~\ref{johnson} to \ref{sdss}. 

\section{The Catalogs}

Two catalogs are produced containing 4413 and 5555 galaxies respectively for 
Abell~1367 and Coma. Both are available electronically from the CDS site.
A one-page sample of the catalogs is given in appendix~A. The catalogs are arranged as follows:\\
(1): Name of the object.\\
(2)-(7): Right Ascension and Declination (J2000), with an accuracy of $\approx$ 2''.\\
(8): Sloan $r'$-band magnitude following the Kron method.\\
(9): Error in the Sloan $r'$-band magnitude, provided by {\em Sextractor}.\\
(10): Surface brightness within the Sloan $r'$ 25.5 mag~arcsec$^{-2}$ isophote.\\
(11): Error in the surface brightness within the Sloan $r'$ 25.5 mag~arcsec$^{-2}$
isophote estimated as the inverse of the signal to noise of the
galaxies within the area enclosed by the 25.5 mag~arcsec$^{-2}$ isophote.\\
(12): Area of the $r'$ 25.5 mag~arcsec$^{-2}$ isophote in arcsec$^{2}$.\\
(13): Number in the catalogs Godwin \& Peach (1982, for Abell~1367) and
Godwin et al. (1983, for Coma). 0 indicates galaxies without an identification
in the Godwin's catalogs.\\
(14): Position angle of the elliptical profile, north to east (in
degrees).\\
(15): Ellipticity ($1 - b/a$, where $a$ and $b$ are respectively the major
and minor axis of the elliptical profile).\\
(16): Star-Galaxy separator (Close to 0 means galaxy-like, close to
1 means star-like).\\

\begin{figure*}[t]
   \centering
   \includegraphics[width=8.5cm]{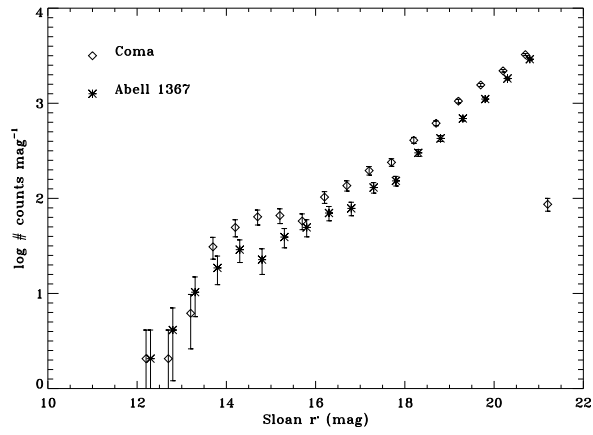}
      \caption{
Galaxy counts in the Sloan $r'$-band for Abell~1367 (asterisks) and Coma
(open diamonds). Error bars include Poisson errors. 
The abscissas have been
shifted $+0.05$ and $-0.05$ magnitudes respectively for Abell~1367 and
Coma, avoiding superposition of points.
}
         \label{histo_rmag}
   \end{figure*}

Figure~\ref{histo_rmag} shows the Sloan $r'$-band total galaxy counts for both
clusters. Error bars contain only the poissonian term. As can
be seen from the figure, the limiting magnitude for both clusters is around
21~mag. However, given the shorter exposure time for one of the fields
in Coma, we adopt a conservative 20.5~mag completeness limit of our survey.

\section{The Luminosity Functions}

\begin{figure}[b]
   \centering
   \includegraphics[width=8.5cm]{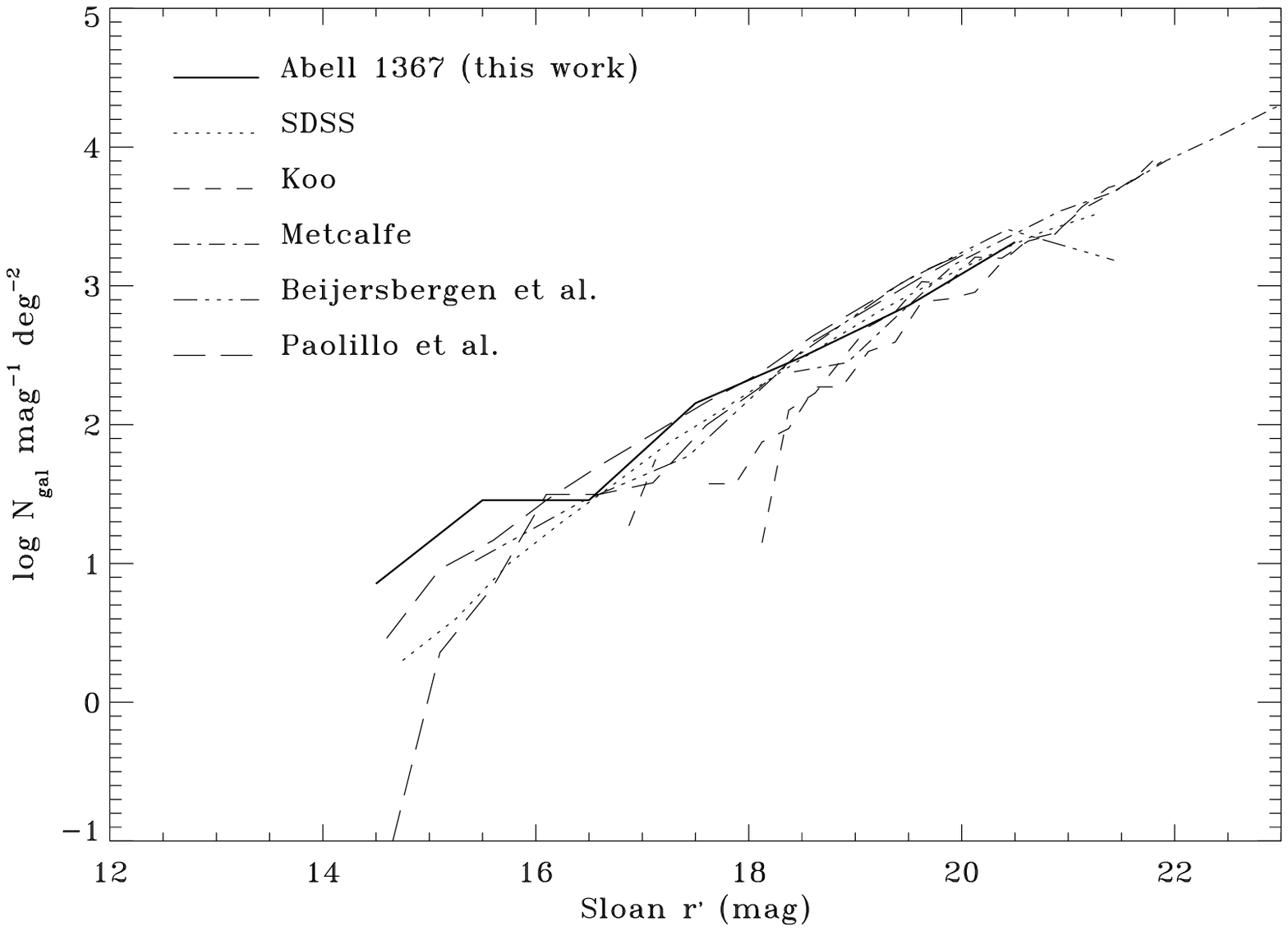}
   \includegraphics[width=8.5cm]{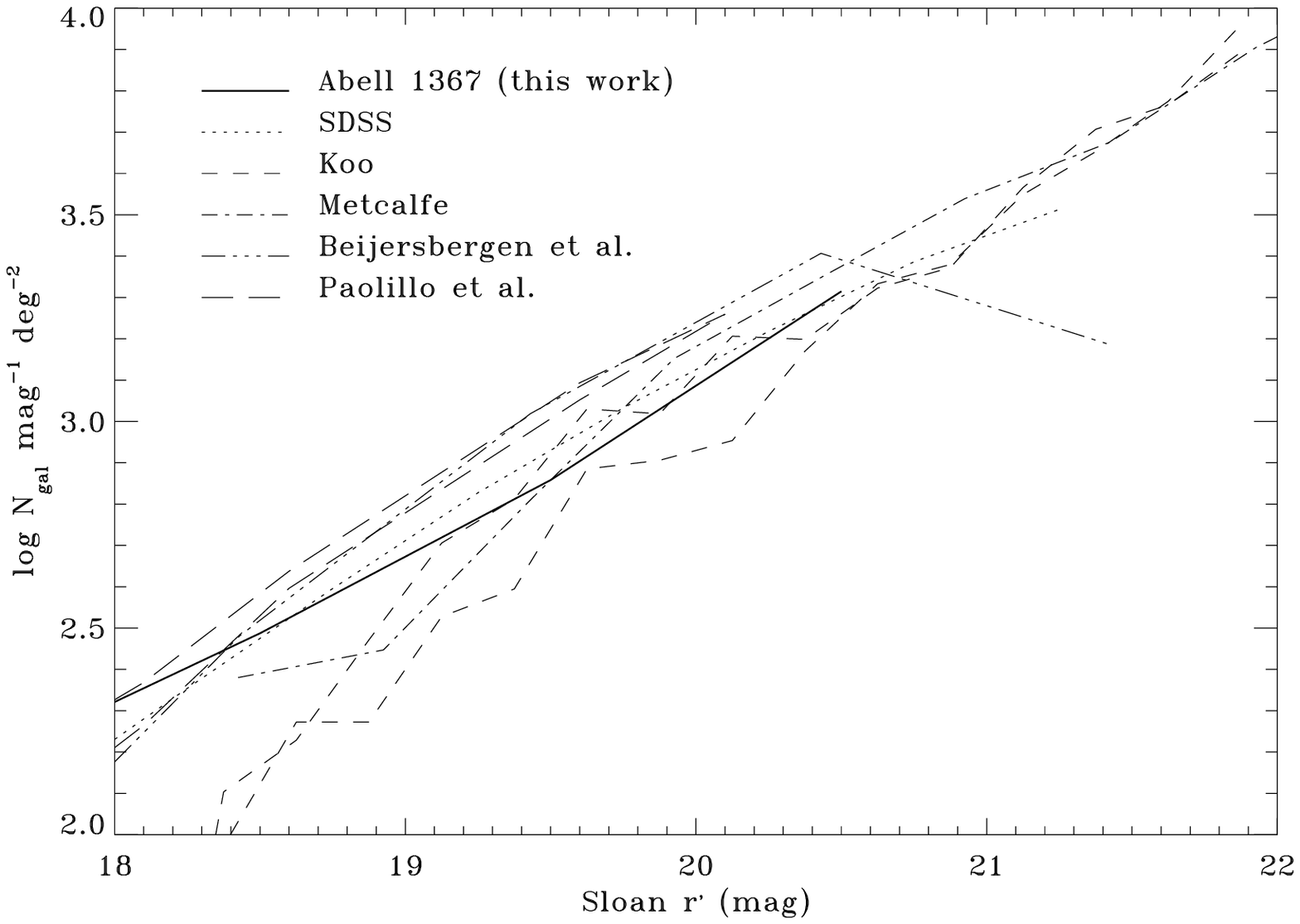}
      \caption{
{\bf Left panel:} Sloan $r'$-band galaxy counts from several sources extracted from
the literature. {\bf Right panel:} Same as left panel zoomed at the
faint magnitude end of the distribution.
}
         \label{check_counts}
   \end{figure}

The membership of galaxies to the two studied clusters is known from complete
redshift measurements only down to 15.5 (Abell~1367) and 16.5 (Coma) magnitudes.
Thus, to construct the $r'$-band LFs 
some statistical subtraction is necessary in order to decontaminate 
our catalogs from background and foreground galaxies. 
Since the aim of our observations was to construct the H$\alpha$ LF 
and not the $r'$ LF, we have not taken observations of a reference field 
from which the galaxy counts could be estimated. Thus, to perform the 
statistical subtraction of background galaxies we must rely on galaxy 
counts taken from the literature.
Various sets of galaxy counts from different sources in the literature 
exist for similar effective wavelengths. 
Figure~\ref{check_counts} shows galaxy counts transformed to the Sloan
$r'$-band system for several sources in the literature (Koo 1986; Metcalfe et
al. 1991; Yasuda et al. 2001; Paolillo et al. 2001; Beijersbergen et
al. 2002). 
A large dispersion exists between the
different galaxy counts sets. We also added for comparison the galaxy
counts from the most external fields of Abell~1367 (i.e. detectors \#2
and \#3 of exposure~1; see Figure~2, left panel in Iglesias-P\'{a}ramo
et al. 2002 for more clarity). Given that these two exposures are about
$1\degr$ far from the cluster center, the contamination due to the
cluster galaxies should be almost negligible. Thus, the counts from
these frames should be reasonably similar to the true background galaxy
counts. As it is apparent in the zoomed right panel of
Figure~\ref{check_counts}, the galaxy counts from the SDSS match
reasonably well our external counts for Abell~1367 towards faint
magnitudes. 
We select this set of galaxy counts in order to decontaminate our
cluster counts for the background and foreground population. We stress
the fact that the results obtained for the LF will depend on the
background counts used to decontaminate the total counts. 
The uncertainty in the galaxy counts due to the cosmic variance of the
background is taken into account in the error budget as it is
explained below.

There is a further concern about the subtraction of the background
counts related to the fact that both the Abell~1367 and Coma clusters
belong to the Great Wall (see Ramella et
al. 1992). Whether the Great Wall should be considered as a background
source for the two clusters is a matter of debate. Gavazzi et
al. (1995) applied a caustic model to determine the membership to the
clusters. As can be seen from their Figure~7, almost
all galaxies with radial velocities within 3$\sigma$ of the average
velocity of any of the two clusters and within the range of projected radial
distances covered by our survey are
considered as members of the clusters. This suggests that the contamination 
by supercluster members is non important at least
within the area covered by our data. Thus, we decided not to apply any
correction due to supercluster population. We remark however that the
analysis by G95 was based on galaxies from the Zwicky catalog, and
nothing can be said about the dwarf galaxies.

A further point concerning the construction of a proper LF is the normalization 
of the galaxy counts to the same area. The total area covered by our mosaic of 
detectors is 1.07 and 1.03 $\degr^{2}$ for Abell~1367 and Coma respectively. 
However, after correction for the area lost because of the presence of strongly 
saturated stars, the gaps between chips and the vignetting at the upper left 
corner of detector \#3, the effective covered area is 0.97~$\degr^{2}$ for both clusters.

In order to account for all the possible sources of error (see
Huang et al. 1997), we included
the contributions from the cosmic variance of the background counts
(this contribution was added twice, to the cluster counts and to the
background counts as suggested by Andreon \& Cuillandre 2002), the
contribution corresponding to the photometric error of the zero point
and the poissonian term.

After subtraction of the background galaxy counts, we fitted the resultant 
points with the Schechter functional form (Schechter 1976):
\begin{equation}
\phi(m_{r'}) = \phi^{*} \times [10^{0.4(m^{*} - m_{r'})}]^{\alpha + 1} 
e^{-10^{0.4(m^{*} - m_{r'})}}
\label{schechter}
\end{equation}

The fitting procedure used minimizes the $\chi^{2}$ and takes 
into account the errors
and assigns to each bin a statistical weight equal to $1/\sigma_{i}^{2}$,
where $\sigma_{i}$ is the error term corresponding to bin $i$.

The value of the parameter
$\alpha$ is of special relevance for our study because it accounts for the relative dwarf-to-giant
populations of galaxies in the clusters.

\subsection{Coma: Comparison with Previous Studies\label{coma_lf_s}}

\begin{figure}[t]
   \centering
   \includegraphics[width=7.5cm]{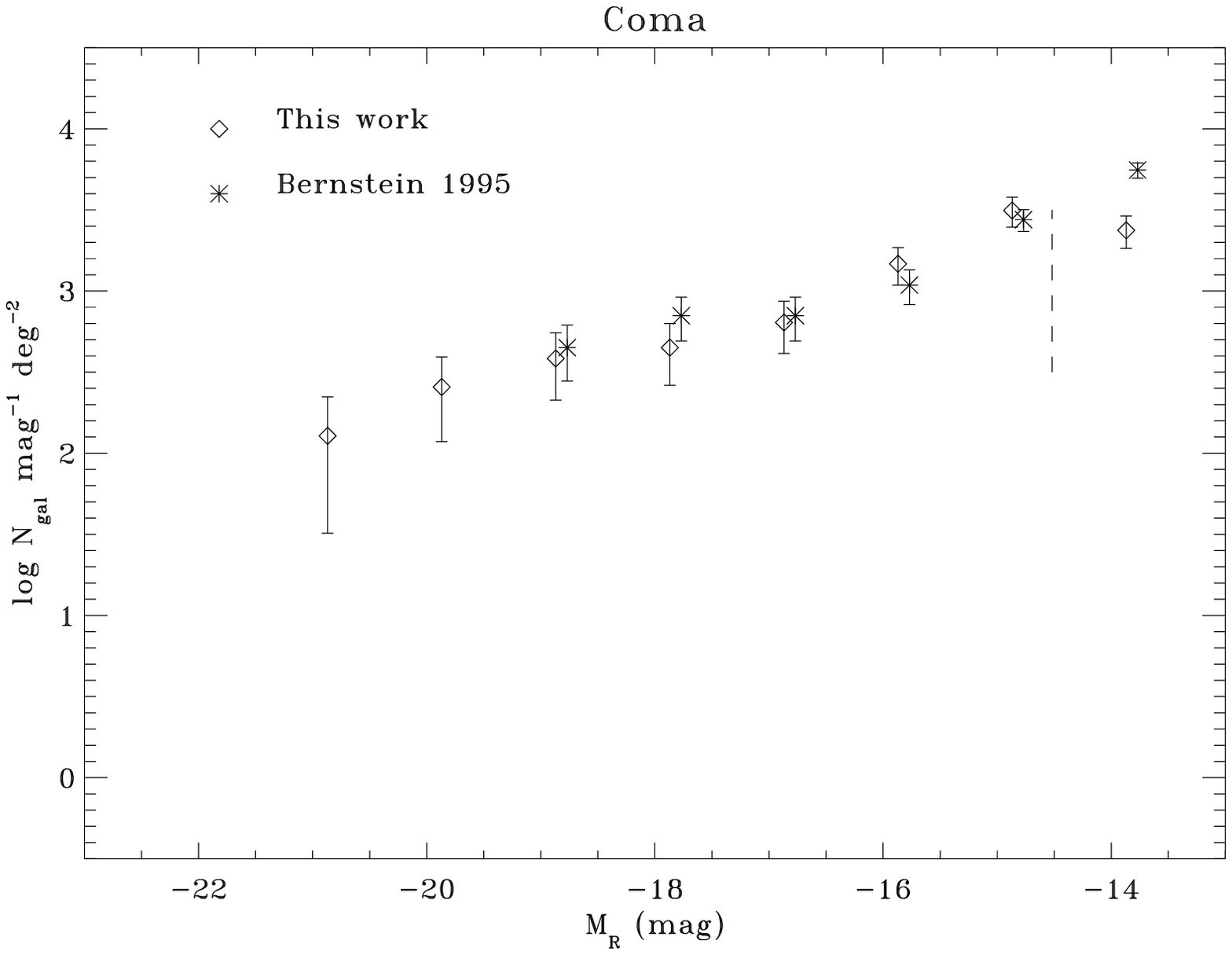}
   \includegraphics[width=7.5cm]{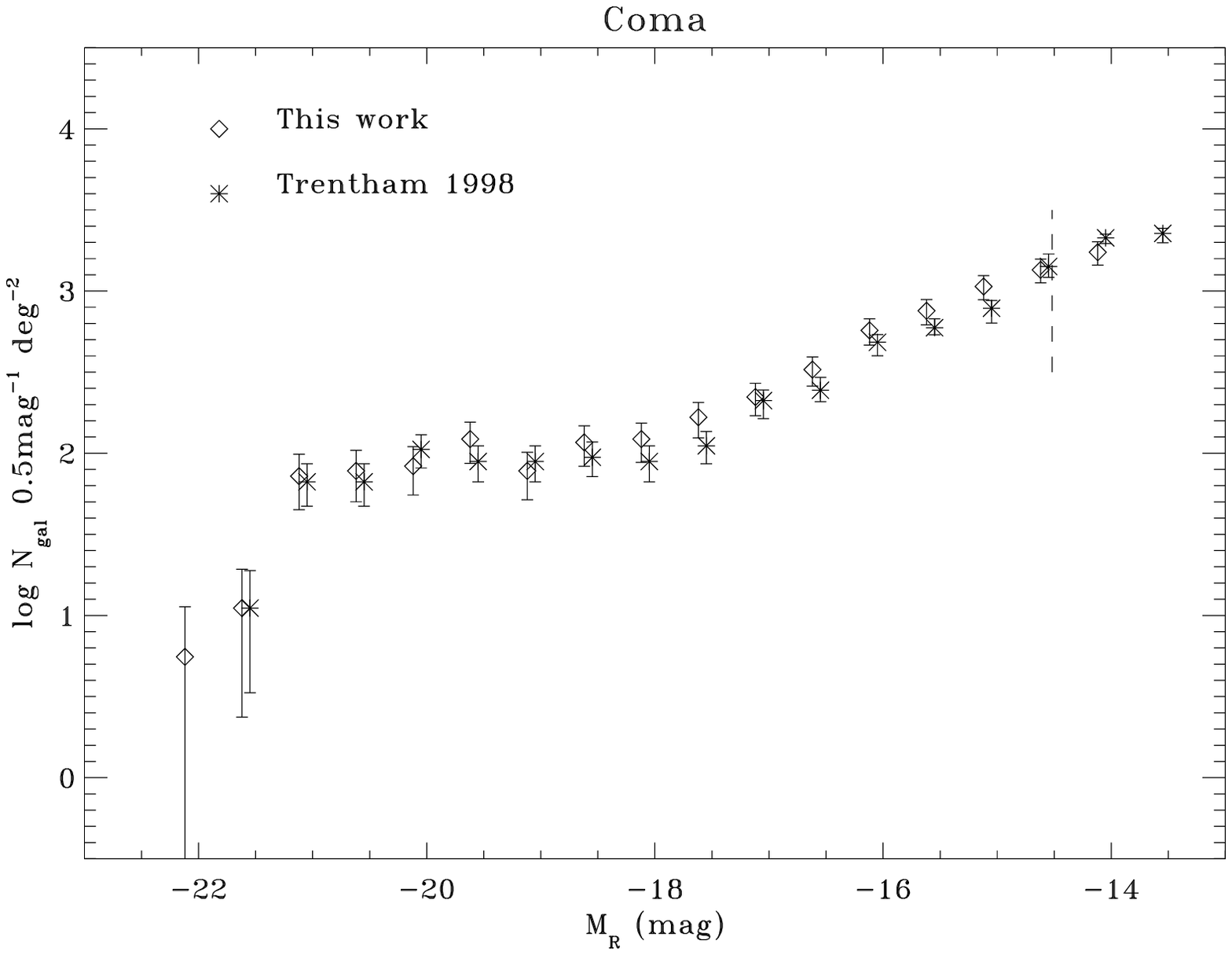}
   \includegraphics[width=7.5cm]{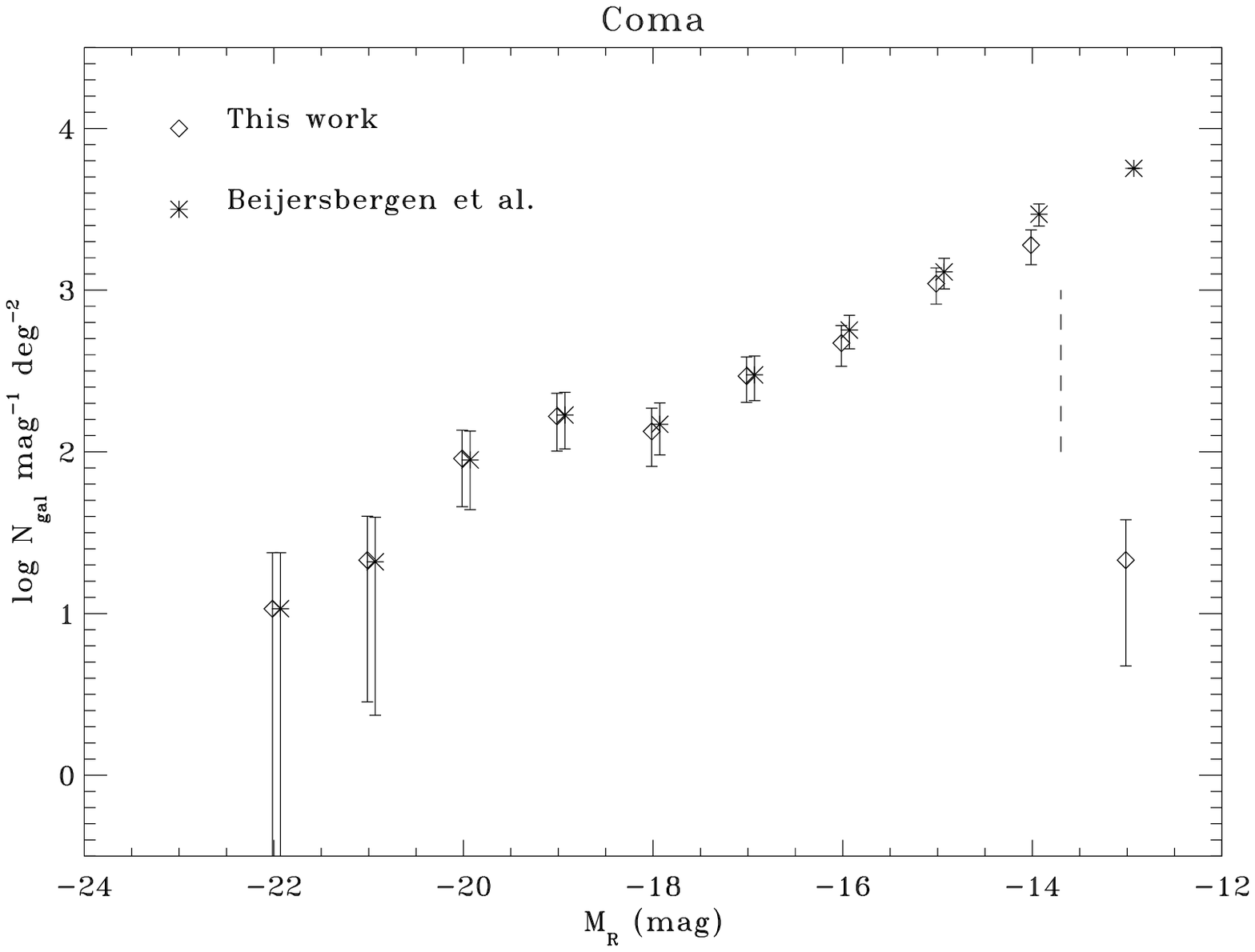}
   \includegraphics[width=7.5cm]{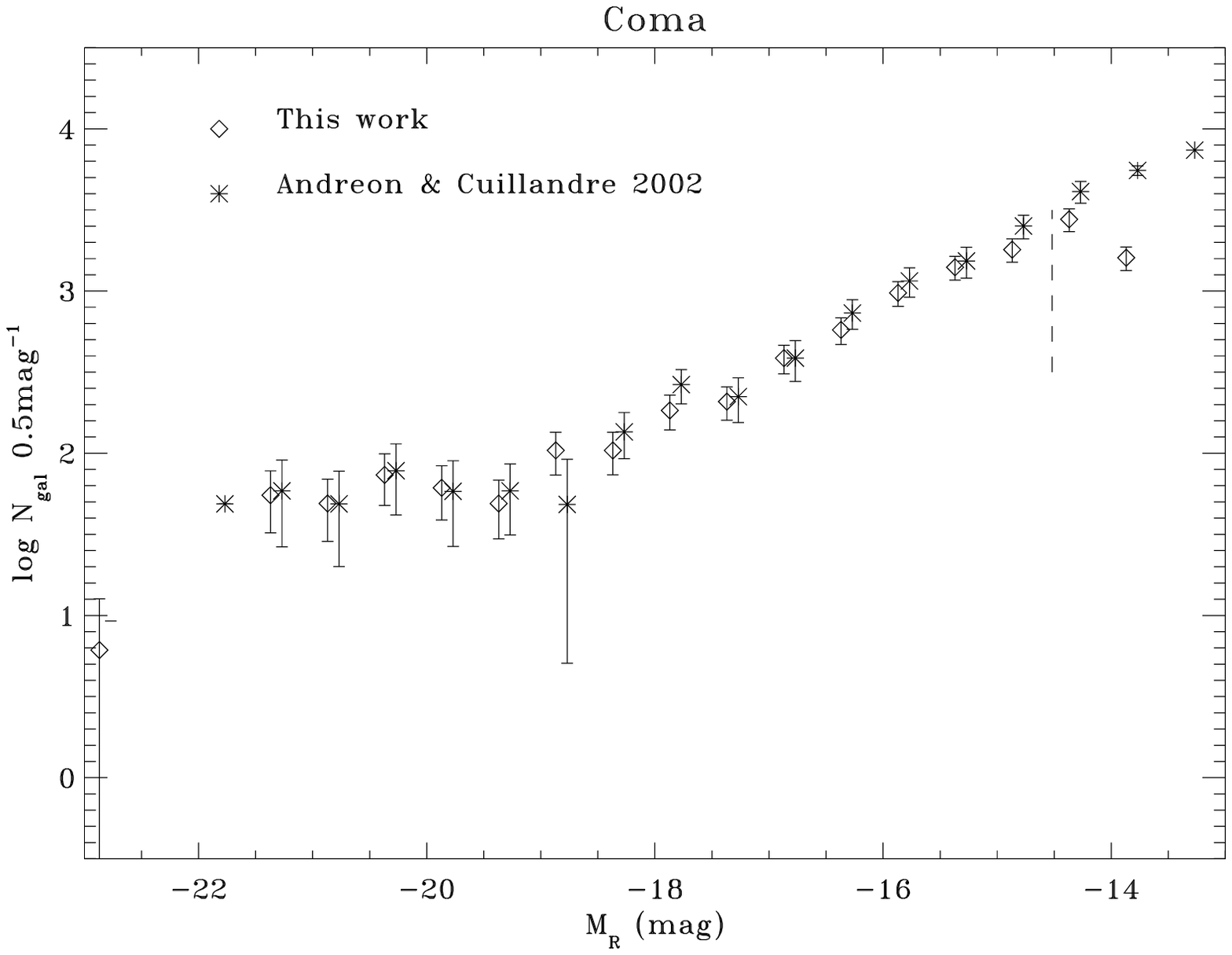}
      \caption{
Comparison of our galaxy counts prior to background subtraction
for Coma with those of Bernstein et
al. (1995), Trentham et al. (1998), Beijersbergen et al. (2002) and
Andreon \& Cuillandre (2002). At each plot, both sets of counts are
refereed to the same region of the sky. The dashed vertical line
indicates our adopted limiting magnitude in the $R_{C}$-band. The
abscissas corresponding to our data and other author's are shifted by $+0.05$ and
$-0.05$~mag respectively for clarity.
}
         \label{compa_bern}
   \end{figure}

As a starting point, we will compare the total counts in the Coma
cluster with others already existing in the literature. 
For this purpose we selected the
compilations by Bernstein et al. (1995), Trentham (1998),
Beijersbergen et al. (2002) and Andreon \& Cuillandre (2002). All these
works present $R_{C}$-band data on the Coma cluster over regions of the
sky totally or partially covered by our survey. The comparison between
our total counts and the ones mentioned above, restricted to a common
region, are presented in Figure~\ref{compa_bern}. Our limiting magnitude, 
$M_{r'} = -14.32$, 
is marked with a vertical dashed line.

Bernstein et al. (1995) obtained very deep $R_{C}$-band CCD imaging of a
small field close to the X-ray center of the cluster, covering an
area of 7.5' square (covered by our survey). 
The two sets of data are consistent within 1$\sigma$
up to our limiting magnitude.

Trentham (1998) obtained data in a small region
(approximately 0.18~$\degr^{2}$, covered by our survey) in the $R_{C}$-band. 
The agreement between the two
sets of data is very good. The slight (within 1$\sigma$) discordance
at $M_{R} \approx -15$~mag could be due to the different method used to extract the
objects: {\em Sextractor} in this work and FOCAS in Trentham's paper.

Beijersbergen et al. (2002) presented data taken with the WFC at 
the INT as ours. Although the area covered by their survey is larger than ours,
they also presented the LF for the inner region of the cluster
which was also covered by us. All points are consistent within 1$\sigma$
in the range of completeness of our data. 

We finally compare our data with those of Andreon \& Cuillandre
(2002). These authors present a very deep survey of the central part of 
the Coma cluster in three bands: $B$, $V$ and
$R$. In this case, the region of the sky covered by these authors did
not exactly match with ours, but the two areas overlap by 90\% of
their total surveyed area.
The data of Andreon \& Cuillandre are normalized to ours at $R_{C} = 18$.
Once again, the data are consistent within our magnitude range of completeness.

The comparison with four independent sources in the literature
shows that our data are consistent within 1$\sigma$ with all of them.
Any residual difference in the derived LF should depend purely on the
adopted background counts, on the effective surveyed area and on the
range of magnitudes over which the LF is computed.

\begin{figure}[t]
   \centering
   \includegraphics[width=8.5cm]{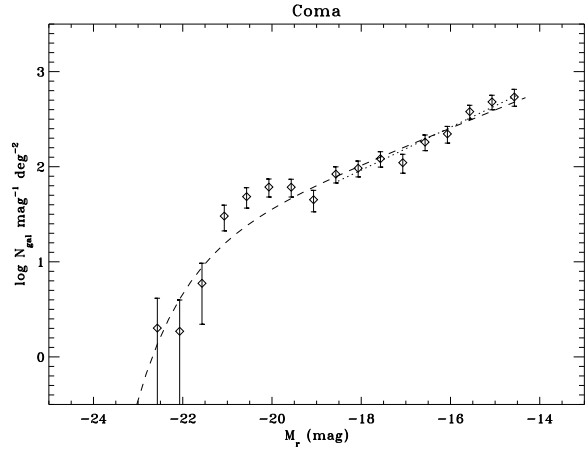}
      \caption{
Sloan $r'$-band LF for Coma in the range of completeness of the
data ($M_{r'} \leq -14.32$). Error bars include the cosmic variance, the zero point and
the poissonian error terms for both the total counts and the background
counts. The best fitting Schechter function is indicated with a
dashed line. The best fitting exponential function in the
range $-19 \leq M_{r'} \leq -14.5$ is indicated with a dotted line.
}
         \label{lf_coma}
   \end{figure}

Figure~\ref{lf_coma} shows our LF for the total area 
covered in the Coma cluster for $r' =
20.5$~mag, assuming the background counts selected in the previous
section. 
The values of the best fitting Schechter parameters
are $\phi^{*} = 21.83 \pm 7.36$, $\alpha = -1.47_{-0.09}^{+0.08}$, 
and $M^{*}_{r'} = -21.63_{-0.57}^{+0.46}$ with $\chi_{\nu}^{2} = 1.69$. 
As previously claimed, we find 
that a Schechter function is not the best representation of the Coma LF in the 
area covered by our survey. Most of the points show deviations by more than
1$\sigma$ from the best fitting function. The distribution
shows two local minima around $M_{r'} \approx -19$ and $-17$
which significantly deviate from the fit at the 1$\sigma$ level. There is
also a maximum centered at $M_{r'} \approx -20.5$ with four points not
fitting the Schechter fit.

In order to make a better estimate of the faint end slope of the LF, 
we fit our data to an exponential function of the type $\approx
10^{km}$ where $m$ is the absolute magnitude and the constant $k$ is
related to the $\alpha$ parameter of the Schechter function by the relation:
\begin{equation}
\alpha = - (k/0.4 + 1)
\end{equation}
The best fitting is shown in Figure~\ref{lf_coma} with a dotted line. 
The best fitting slope is $k = 0.22 \pm 0.03$
with $\chi_{\nu}^{2} = 0.77$ in the magnitude range $-19 \leq M_{r'} \leq
-14.5$ (corresponding to $\alpha = -1.55)$.

\subsection{Abell~1367}

\begin{figure}[t]
   \centering
   \includegraphics[width=8.5cm]{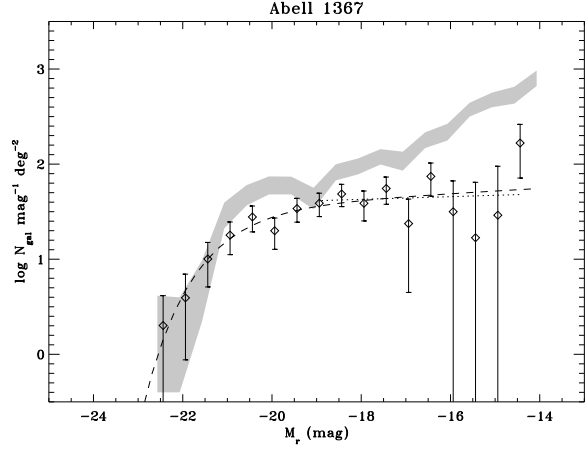}
      \caption{
Sloan $r'$-band LF for Abell~1367 in the range of completeness of the
data ($M_{r'} \leq -14.19$). Error bars include the same terms as in
Figure~\ref{lf_coma}. The dashed line indicates the best fitting to a
Schechter function. The dotted line corresponds to the best fitting to
an exponential function in the range $-19 \leq M_{r'} \leq -14.5$. 
The shaded region is the Coma LF within its
1$\sigma$ error bars.
}
         \label{lf_ab}
   \end{figure}

Figure~\ref{lf_ab} shows our LF for the total area 
covered in Abell~1367 in the magnitude range of completeness 
($M_{r'} \leq -14.19$).
The best fitting Schechter parameters
are $\phi^{*} = 35.66 \pm 12.40$, $\alpha = -1.07_{-0.16}^{+0.20}$, 
and $M^{*}_{r'} = -21.20_{-0.63}^{+0.60}$ with $\chi_{\nu}^{2} = 0.53$.
In this case, the Schechter function provides a good fitting of the
data except for 3 points which do not fit the model within 1$\sigma$.
The errors
become very large for $M_{r'} \geq -17$. 
We also fit an exponential function 
over the same range of magnitudes as for the Coma LF and obtain a
value for the slope of $k = 0.02 \pm 0.06$, corresponding to $\alpha
= -1.05$, with $\chi_{\nu}^{2} = 0.66$. 

The comparison of the LFs of Coma and Abell~1367 is shown in
Figure~\ref{lf_ab} (the shaded band corresponding to the uncertainty
region of Coma LF). As can be seen from the plot, at most of the
magnitudes, LFs of both clusters do not coincide at the 1$\sigma$
level. This effect is mostly due to the different richness of both
clusters. However, the shapes of the LFs also show appreciable
differences:
The LF of Abell~1367 does not show the bump at
$M_{r'} \approx -20$ exhibited by the Coma LF. The steep rise of the
slope of the Abell~1367 LF in the interval $-15.5 \leq r' \leq -14.5$
cannot be trusted due to large statistical uncertainties.

\begin{figure}[t]
   \centering
   \includegraphics[width=8.5cm]{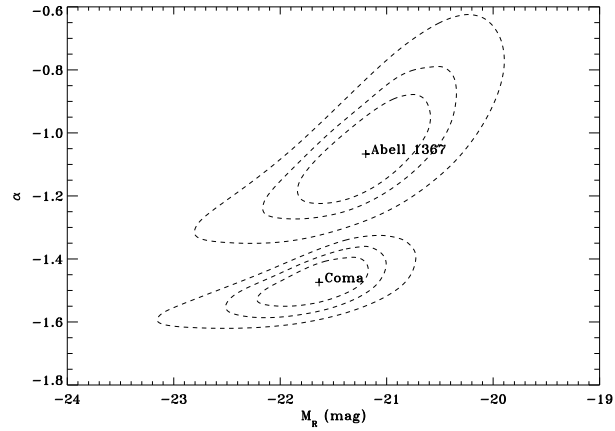}
      \caption{
1, 2 and 3$\sigma$ confidence contours for the best fitting Schechter function
parameters of Abell~1367 and Coma.
}
         \label{ellipse_lf}
   \end{figure}

Figure~\ref{ellipse_lf} shows the 1, 2 and 3$\sigma$ confidence contours for the
best fitting Schechter function parameters of both clusters. 
Although the errors in the determination of the parameters of the
Abell~1367 LF are large, even the 3$\sigma$ contours do not cross each
other, supporting the conclusion that the slopes of the two clusters are
significantly different at this level. 

The ratio of dwarf-to-giant galaxies is considerably higher in 
Coma than in Abell~1367. This result, together with the fact that the
H$\alpha$ LFs of both clusters are fairly similar (Iglesias-P\'{a}ramo et
al. 2002), indicates that an important population of non star-forming 
dwarf galaxies present in Coma is absent in Abell~1367.

\subsection{The Influence of the Cluster Environment on the LFs of Coma
and Abell~1367}

Beijersbergen et al. (2002) found that the faint end of the LF of
the Coma cluster steepens as we move outwards in the cluster. We
repeat a similar exercise for the two studied clusters. In order to
minimize errors due to the limited statistics (mostly severe for the
outer parts of Abell~1367), we select only two regions of each
cluster: an inner one of radius $0.4\degr$ and an external one
with projected radius larger than 0.5$\degr$. 
Galaxies belonging to the annulus $0.4\degr < R < 0.5\degr$ are
excluded from this analysis. 
The cluster centers were assumed coincident with the peak of the extended
X-ray sources (Donnelly et al. 1998 for Abell~1367; White et al. 1993
for Coma).
We use the same radius for both clusters because not only are they at 
approximately the same distance, but also the estimates of their physical sizes 
coincide (see Girardi et al. 1995). The virial radii were estimated as 0.5$\degr$ 
and 0.4$\degr$ and the core radii
as 0.12$\degr$ and 0.08$\degr$ respectively for Abell~1367 and Coma.

\begin{figure*}[t]
   \centering
   \includegraphics[width=8.5cm]{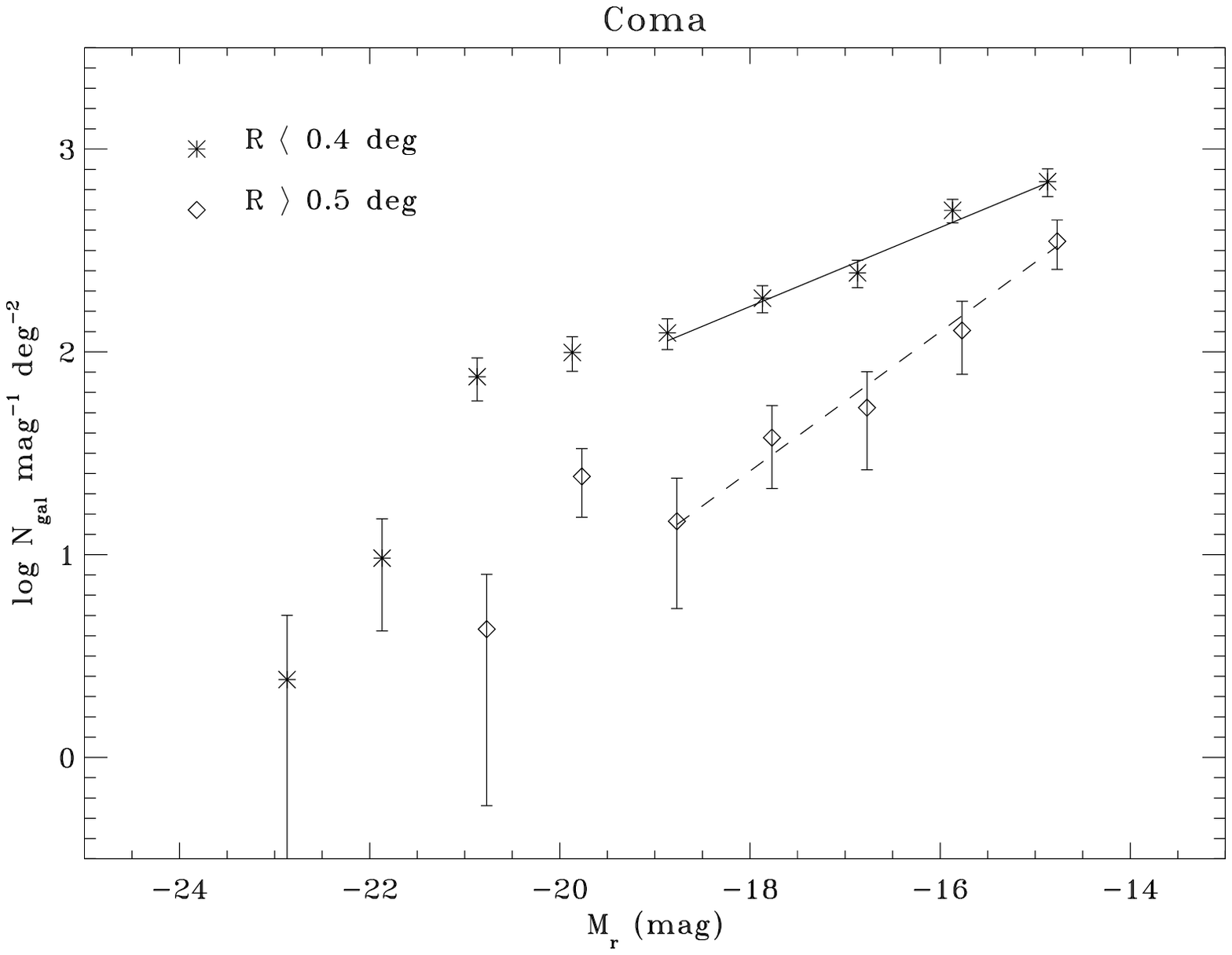}
   \includegraphics[width=8.5cm]{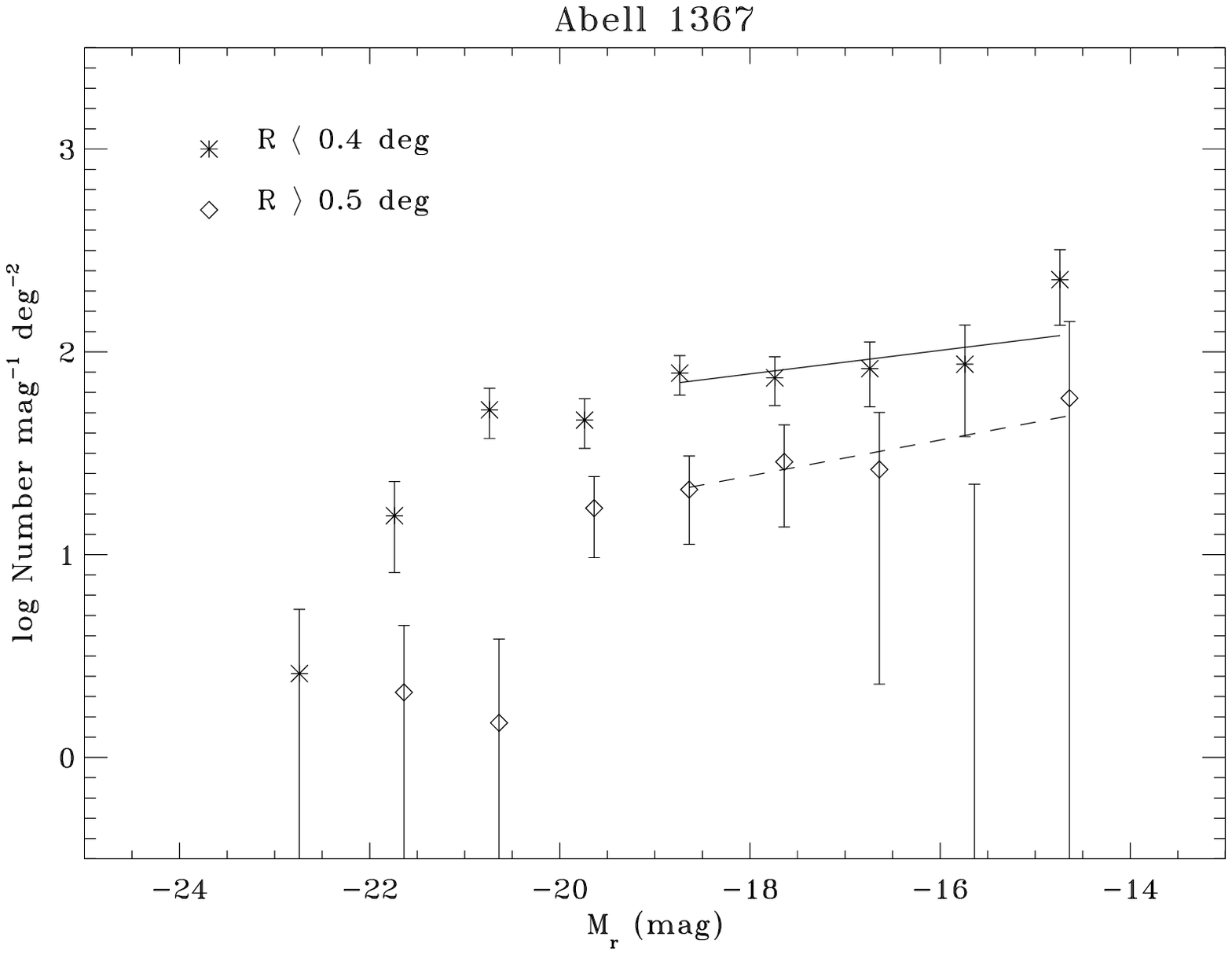}
      \caption{
Sloan $r'$-band LFs for the inner and outer regions of Abell~1367 (left plot) and Coma
(right plot). 
}
         \label{env_ab}
   \end{figure*}

Figure~\ref{env_ab} shows the LFs of the two clusters restricted to the two 
regions mentioned above. The data are binned by 1~mag in order to
increase the statistics, and fitted to exponential functions.
As can be seen in the left panel of Figure~\ref{env_ab} the slopes of
the LFs of the two regions of Coma are clearly different from each
other. 
The best
fit values for the slopes over the range $-19 \leq M_{r} \leq -14.5$
are $k_{in} = 0.19 \pm 0.03$ and $k_{out} = 0.34 \pm 0.07$  for the inner and outer regions 
respectively. 

For Abell~1367 the restricted LFs are shown in the right panel of the same
figure. The cluster counts become
negative at $M_{r'} \approx -15.5$ for the external annulus and the
error bars are very large fainter than $M_{r'} =
-16.5$ affecting the reliability of this comparison.
The slopes obtained from the best fitting to the exponential function (in the
same magnitude range as for Coma) are
$k_{in} = 0.06 \pm 0.06$ and $k_{out} = 0.09 \pm 0.11$ for 
the inner and outer regions respectively, thus showing consistency.
However, we stress that the slope for the outer annulus was
computed rejecting the negative point at $M_{r'} \approx -15.5$, and
that the real uncertainty could be even larger than the one obtained from the
least squared fitting.

\section{Summary and Conclusions}

We present new deep catalogs containing positions and $r'$-band photometry of
galaxies in the central $1\degr \times 1\degr$ of the nearby clusters Abell~1367 and Coma. 
These catalogs are used to determine the SDSS $r'$-band LFs of both
clusters by subtracting the Yasuda et al. (2001) galaxy counts from
our cluster counts. 
The faint-end slope of the Coma LF is $\alpha = -1.47_{-0.09}^{+0.08}$
whereas that of the Abell~1367 LF
is shallower, with $\alpha =
-1.07_{-0.16}^{+0.20}$. This difference is found significant at
the 3$\sigma$ level.
Given that the observations of both clusters were
obtained in homogeneous conditions, we argue that these differences
are not due to instrumental or data handling biases, but they are
intrinsic to the clusters. We also stress that the differences
are not due to the background counts since the best set of galaxy
counts was used to decontaminate the counts of both clusters.

The LF parameters strongly depend on the surveyed region, 
on the background counts used to decontaminate the cluster counts
and on the magnitude range over which the fitting is performed.
Other determinations of the LFs of the same clusters (in
section~6.1 we checked that our total Coma counts are
consistent with several sources in the literature), is valid only 
with those
covering a similar area and with substantial overlapping.

Concerning the comparison with the LF of field galaxies, the values of
$\alpha$ derived for Coma and Abell~1367 are consistent with those of
the field, which show a larger spread of parameters: Lin et al. (1996) found $\alpha =
-0.70 \pm 0.05$ ($M_{R} < -17.5$), Geller et al. (1997) found $\alpha = 
-1.17 \pm 0.19$ ($M_{R} < -16$) and Blanton et al. (2001) reported
$\alpha = -1.20 \pm 0.03$ ($M_{r'} < -16$).

The slope of the LF of Coma is found steeper towards the
cluster outskirts within 1$\sigma$ statistical significance. No such
trend is observed in Abell~1367.
This means that the bright-to-faint galaxy ratio in Coma decreases as we 
move outwards the cluster. 
The observed increase of $\alpha$ in the cluster's outskirts could be
explained both by an increase of the dwarf population or by a decrease
of the giant population with respect to the cluster center.
It would be interesting to find out at which clustercentric distance
the cluster LF would approach the one of the field. 
Once again Coma and Abell~1367 would be
the ideal clusters to do this test: they are both embedded 
in the Great Wall (Geller \& Huchra 1990) located at the 
same distance of $\sim$ 6500-7000
km~s$^{-1}$. A survey of some square degrees in between the two
clusters would suffice to map the variations of the LF with increasing
clustercentric distances. Given the low density of supercluster
galaxies, a deep redshift survey is however needed.

\begin{acknowledgements}
We thank S.Andreon for his interesting comments and suggestions.
This research has made use of the NASA/IPAC Extragalactic Database (NED) which
is operated by the Jet Propulsion Laboratory, California Institute of
Technology, under contract with the National Aeronautics and Space
Administration. The INT is operated on the island of La Palma 
by the ING group, in the Spanish Observatorio del 
Roque de Los Muchachos of the Instituto de Astrof\'{\i}sica 
de Canarias. JIP acknowledges the Fifth Framework
Program of the EU for a Marie Curie Postdoctoral Fellowship.
\end{acknowledgements}

\newpage

\onecolumn

   \begin{table}[b]
      \caption[]{Percentage of galaxies for which $|r'_{Aper} - r'_{Sex}| > 0.2$,
for different magnitude intervals.}
         \label{close}
     $$ 
         \begin{tabular}{lcc}
            \hline
            \noalign{\smallskip}
Magnitude Interval & Fraction (\%) \\
            \noalign{\smallskip}
            \hline
            \noalign{\smallskip}
$15.5 \leq r' < 17.5$ & 0.9 \\
$17.5 \leq r' < 19.5$ & 2.9 \\
$19.5 \leq r' < 21.0$ & 8.6 \\
            \noalign{\smallskip}
            \hline
         \end{tabular}
     $$ 
   \end{table}

   \begin{table*}
      \caption[]{Number counts and errors for both clusters and the
background counts in the SDSS
photometric system (Sloan $r'$-band magnitudes). To be published electronically.}
         \label{cuentas}
     $$ 
     $$ 
   \end{table*}

\newpage

\clearpage

\newpage

\clearpage

\appendix

\clearpage

\section{Overview of the Catalogs}

\scriptsize

{\tt

\begin{tabular}{lrrrrrrrrrrrrrrr}
\hline
Name & hh & mm & ss & dd & md & sd & $r'$ & $\Delta r'$  & $\mu_{25.5}$ &
$\Delta \mu_{25.5}$ & Area & God & $\theta$ & Ellip. & S/G \\
 & \multicolumn{3}{c}{(J2000.)} & \multicolumn{3}{c}{(J2000.)} & 
\multicolumn{2}{c}{(mag)} & \multicolumn{2}{c}{(mag\,arcsec$^{-2}$)} & 
arcsec$^{2}$ & & (deg.) & & \\
\hline
113957+194339 & 11 & 39 & 57.8 & 19 & 43 & 39 & 20.68 &  0.02 & 23.88 &  0.02 &   17.96 &    0 & 174.59 &  0.44 &  0.53 \\
113958+195514 & 11 & 39 & 58.4 & 19 & 55 & 14 & 20.08 &  0.01 & 23.80 &  0.02 &   27.72 &    0 & 157.09 &  0.07 &  0.07 \\
113958+194515 & 11 & 39 & 58.5 & 19 & 45 & 15 & 19.88 &  0.01 & 23.64 &  0.02 &   32.16 &    0 &  69.50 &  0.15 &  0.03 \\
113958+194329 & 11 & 39 & 58.7 & 19 & 43 & 29 & 20.84 &  0.02 & 23.84 &  0.02 &   14.30 &    0 &  72.89 &  0.20 &  0.47 \\
113958+194333 & 11 & 39 & 58.8 & 19 & 43 & 33 & 19.32 &  0.01 & 23.48 &  0.02 &   41.14 &    0 & 162.66 &  0.05 &  0.03 \\
113958+194827 & 11 & 39 & 58.8 & 19 & 48 & 27 & 17.79 &  0.00 & 22.99 &  0.01 &  111.89 &    0 & 139.41 &  0.44 &  0.03 \\
113958+194702 & 11 & 39 & 58.8 & 19 & 47 &  2 & 20.87 &  0.02 & 24.06 &  0.03 &   17.52 &    0 &  55.84 &  0.33 &  0.01 \\
113958+200102 & 11 & 39 & 58.9 & 20 &  1 &  2 & 19.26 &  0.01 & 23.46 &  0.02 &   43.25 &    0 &  19.05 &  0.25 &  0.03 \\
113958+195339 & 11 & 39 & 59.0 & 19 & 53 & 39 & 19.72 &  0.01 & 23.43 &  0.02 &   27.39 &    0 &  54.46 &  0.05 &  0.29 \\
113959+194251 & 11 & 39 & 59.1 & 19 & 42 & 51 & 20.21 &  0.02 & 24.00 &  0.02 &   26.17 &    0 &  62.95 &  0.41 &  0.16 \\
113959+195403 & 11 & 39 & 59.1 & 19 & 54 &  3 & 20.58 &  0.02 & 24.14 &  0.03 &   23.40 &    0 & 164.80 &  0.18 &  0.02 \\
113959+194743 & 11 & 39 & 59.3 & 19 & 47 & 43 & 20.94 &  0.02 & 24.04 &  0.02 &   13.42 &    0 & 104.87 &  0.13 &  0.63 \\
113959+194703 & 11 & 39 & 59.3 & 19 & 47 &  3 & 20.52 &  0.02 & 24.06 &  0.03 &   22.40 &    0 & 137.74 &  0.25 &  0.03 \\
113959+194700 & 11 & 39 & 59.6 & 19 & 47 &  0 & 20.43 &  0.01 & 23.97 &  0.02 &   22.51 &    0 & 116.33 &  0.20 &  0.03 \\
113959+194948 & 11 & 39 & 59.8 & 19 & 49 & 48 & 20.24 &  0.01 & 23.95 &  0.02 &   26.17 &    0 &  26.36 &  0.22 &  0.02 \\
113959+195210 & 11 & 39 & 59.8 & 19 & 52 & 10 & 19.40 &  0.01 & 23.45 &  0.02 &   38.70 &    0 &  95.13 &  0.05 &  0.20 \\
113959+195346 & 11 & 39 & 59.8 & 19 & 53 & 46 & 20.92 &  0.02 & 24.12 &  0.03 &   16.19 &    0 &  25.87 &  0.11 &  0.37 \\
113959+195456 & 11 & 39 & 59.9 & 19 & 54 & 56 & 19.87 &  0.01 & 23.54 &  0.02 &   26.17 &    0 &  70.95 &  0.20 &  0.13 \\
114000+194327 & 11 & 40 &  0.1 & 19 & 43 & 27 & 18.82 &  0.01 & 23.41 &  0.01 &   61.77 &    0 & 118.40 &  0.20 &  0.03 \\
114000+195714 & 11 & 40 &  0.2 & 19 & 57 & 14 & 19.96 &  0.01 & 23.68 &  0.02 &   27.61 &    0 &  27.04 &  0.07 &  0.07 \\
114000+195343 & 11 & 40 &  0.3 & 19 & 53 & 43 & 17.42 &  0.00 & 22.97 &  0.01 &  155.24 &    0 & 164.95 &  0.23 &  0.08 \\
114000+194340 & 11 & 40 &  0.5 & 19 & 43 & 40 & 19.92 &  0.01 & 23.78 &  0.02 &   30.38 &    0 &  38.22 &  0.10 &  0.05 \\
114000+195426 & 11 & 40 &  0.6 & 19 & 54 & 26 & 15.63 &  0.00 & 22.39 &  0.01 &  462.96 &    0 & 152.53 &  0.20 &  0.03 \\
114000+195048 & 11 & 40 &  0.7 & 19 & 50 & 48 & 18.93 &  0.01 & 23.08 &  0.01 &   41.81 &    0 &  57.56 &  0.26 &  0.10 \\
114001+195315 & 11 & 40 &  1.1 & 19 & 53 & 15 & 20.60 &  0.02 & 23.83 &  0.02 &   16.52 &    0 &  29.89 &  0.07 &  0.12 \\
114001+194401 & 11 & 40 &  1.2 & 19 & 44 &  1 & 18.84 &  0.01 & 23.12 &  0.01 &   47.02 &    0 & 102.02 &  0.17 &  0.04 \\
114001+195729 & 11 & 40 &  1.3 & 19 & 57 & 29 & 19.18 &  0.01 & 23.22 &  0.01 &   37.48 &    0 & 109.11 &  0.14 &  0.04 \\
114001+195330 & 11 & 40 &  1.3 & 19 & 53 & 30 & 20.00 &  0.01 & 24.07 &  0.03 &   36.15 &    0 &  80.50 &  0.26 &  0.02 \\
114001+194355 & 11 & 40 &  1.3 & 19 & 43 & 55 & 19.30 &  0.01 & 23.48 &  0.02 &   41.81 &    0 & 149.09 &  0.20 &  0.06 \\
114001+200001 & 11 & 40 &  1.4 & 20 &  0 &  1 & 19.67 &  0.01 & 24.44 &  0.03 &   55.56 &    0 &   6.00 &  0.23 &  0.00 \\
114001+195821 & 11 & 40 &  1.7 & 19 & 58 & 21 & 19.77 &  0.01 & 23.72 &  0.02 &   33.16 &    0 & 114.59 &  0.32 &  0.03 \\
114002+195313 & 11 & 40 &  2.4 & 19 & 53 & 13 & 19.36 &  0.01 & 23.32 &  0.01 &   35.15 &    0 &  86.57 &  0.17 &  0.03 \\
114002+195324 & 11 & 40 &  2.5 & 19 & 53 & 24 & 20.39 &  0.01 & 23.81 &  0.02 &   20.85 &    0 &  58.13 &  0.16 &  0.05 \\
114002+194219 & 11 & 40 &  2.5 & 19 & 42 & 19 & 20.43 &  0.01 & 23.99 &  0.02 &   22.95 &    0 &  26.57 &  0.11 &  0.03 \\
114002+194331 & 11 & 40 &  2.7 & 19 & 43 & 31 & 19.02 &  0.01 & 23.34 &  0.01 &   49.57 &    0 &  49.21 &  0.10 &  0.03 \\
114002+200227 & 11 & 40 &  2.7 & 20 &  2 & 27 & 20.89 &  0.03 & 24.38 &  0.03 &   16.08 &    0 & 125.59 &  0.18 &  0.04 \\
114002+194931 & 11 & 40 &  2.9 & 19 & 49 & 31 & 18.27 &  0.01 & 23.24 &  0.01 &   89.93 &    0 &   9.50 &  0.16 &  0.03 \\
114002+194212 & 11 & 40 &  3.0 & 19 & 42 & 12 & 17.33 &  0.00 & 23.08 &  0.01 &  185.18 &    0 & 163.54 &  0.01 &  0.03 \\
114003+195245 & 11 & 40 &  3.2 & 19 & 52 & 45 & 20.12 &  0.01 & 23.67 &  0.02 &   24.84 &    0 &  70.51 &  0.29 &  0.03 \\
114003+194603 & 11 & 40 &  3.4 & 19 & 46 &  3 & 20.35 &  0.01 & 23.94 &  0.02 &   23.51 &    0 &  54.08 &  0.17 &  0.03 \\
114003+200148 & 11 & 40 &  3.7 & 20 &  1 & 48 & 19.74 &  0.01 & 23.60 &  0.02 &   28.05 &    0 & 123.35 &  0.20 &  0.70 \\
114003+194354 & 11 & 40 &  3.7 & 19 & 43 & 54 & 19.80 &  0.01 & 23.71 &  0.02 &   32.82 &    0 &  94.62 &  0.23 &  0.03 \\
114003+200010 & 11 & 40 &  3.9 & 20 &  0 & 10 & 20.12 &  0.01 & 23.94 &  0.02 &   30.83 &    0 &  20.88 &  0.19 &  0.04 \\
114003+195325 & 11 & 40 &  4.0 & 19 & 53 & 25 & 20.65 &  0.02 & 24.34 &  0.03 &   24.51 &    0 &  20.58 &  0.46 &  0.01 \\
114004+195129 & 11 & 40 &  4.1 & 19 & 51 & 29 & 20.12 &  0.01 & 23.73 &  0.02 &   24.84 &    0 &  56.35 &  0.11 &  0.04 \\
114004+195808 & 11 & 40 &  4.3 & 19 & 58 &  8 & 20.59 &  0.02 & 23.92 &  0.02 &   18.19 &    0 &  83.01 &  0.20 &  0.42 \\
114004+194900 & 11 & 40 &  4.4 & 19 & 49 &  0 & 20.54 &  0.02 & 23.81 &  0.02 &   18.07 &    0 &  64.56 &  0.08 &  0.62 \\
114004+194414 & 11 & 40 &  4.5 & 19 & 44 & 14 & 18.83 &  0.01 & 23.07 &  0.01 &   45.80 &    0 &  41.12 &  0.06 &  0.04 \\
114004+195552 & 11 & 40 &  4.7 & 19 & 55 & 52 & 19.73 &  0.01 & 23.59 &  0.02 &   32.60 &    0 &  66.50 &  0.17 &  0.03 \\
114004+194522 & 11 & 40 &  4.8 & 19 & 45 & 22 & 18.99 &  0.01 & 23.66 &  0.02 &   66.31 &    0 &  78.35 &  0.28 &  0.03 \\
114005+195754 & 11 & 40 &  5.2 & 19 & 57 & 54 & 19.77 &  0.01 & 23.49 &  0.02 &   27.83 &    0 & 135.47 &  0.07 &  0.41 \\
114005+194544 & 11 & 40 &  5.6 & 19 & 45 & 44 & 19.73 &  0.01 & 23.70 &  0.02 &   34.71 &    0 &  47.48 &  0.11 &  0.05 \\
114005+200220 & 11 & 40 &  5.6 & 20 &  2 & 20 & 20.31 &  0.02 & 24.03 &  0.03 &   26.84 &    0 &  77.06 &  0.11 &  0.02 \\
114006+200151 & 11 & 40 &  6.1 & 20 &  1 & 51 & 20.34 &  0.02 & 24.19 &  0.03 &   30.05 &    0 &  52.99 &  0.12 &  0.01 \\
114006+200216 & 11 & 40 &  6.2 & 20 &  2 & 16 & 20.61 &  0.02 & 23.89 &  0.02 &   17.96 &    0 &  47.88 &  0.09 &  0.32 \\
\end{tabular}

}

\end{document}